\documentclass[aps,prx,preprintnumbers,amsmath,amssymb,superscriptaddress,twocolumn,10pt, floatfix]{revtex4-2}
\usepackage{lmodern}
\usepackage{graphicx}
\usepackage{dcolumn}
\usepackage{bm, bbm}
\usepackage{array, xspace}
\usepackage[normalem]{ulem}
\usepackage[dvipsnames]{xcolor}
\usepackage{xr-hyper} 
\usepackage[%
    colorlinks=true,
    pdfborder={0 0 0},
    linkcolor=red
]{hyperref}
\usepackage{chngpage, float}
\usepackage{appendix}

\begin{document}
\def\ie{\emph{i.e.,} }
\def\cvs {CsV$_3$Sb$_5$\xspace}
\def\cvsn {CsV$_3$Sb$_{5-x}$Sn$_x$\xspace}
\def\cvsTi {Cs(V$_{1-x}$Ti$_x$)Sb$_{5}$\xspace}
\def\V {$^{51}$V\xspace}
\def\Sb {$^{121}$Sb\xspace}
\def\Rb {$^{87}$Rb\xspace}
\def\usr {µSR\xspace}
\def\isdpi {\mbox{ISD-$\pi$}\xspace}

\title{Observation of ubiquitous charge correlations and hidden quantum critical point in hole-doped kagome superconductors}

\author{\hspace{1mm}Ilija K.~Nikolov}
\affiliation{
Department of Physics, Brown University, Providence, Rhode Island 02912, USA}

\author{\hspace{1mm}Giuseppe Allodi}
\affiliation{
Dipartimento di Scienze Matematiche, Fisiche e Informatiche, Universit\`a di Parma, I-43124 Parma, Italy
}

\author{\hspace{1mm}Adrien Rosuel}
\affiliation{
Department of Physics, Brown University, Providence, Rhode Island 02912, USA}
 \affiliation{
Brown Center for Theoretical Physics and Innovation, BCTPI, Brown University, Providence, RI 02912-1843, USA}

\author{\hspace{1mm}Ginevra Corsale}
\affiliation{
Department of Physics, Brown University, Providence, Rhode Island 02912, USA}
 \affiliation{Dipartimento di Fisica e Astronomia, Universit\`a di Bologna, I-40127 Bologna, Italy}

\author{\hspace{1mm}Anshu Kataria}
\affiliation{
Dipartimento di Scienze Matematiche, Fisiche e Informatiche, Universit\`a di Parma, I-43124 Parma, Italy
}

\author{\hspace{1mm}Pietro Bonf\`a}
\affiliation{
Dipartimento di Fisica, Informatica e Matematica, Universit\`a di Modena e Reggio Emilia, Via Campi 213/a, 41125 Modena, Italy
}

\author{\hspace{1mm}Roberto De Renzi}
\affiliation{
Dipartimento di Scienze Matematiche, Fisiche e Informatiche, Universit\`a di Parma, I-43124 Parma, Italy
}

\author{\hspace{1mm}Andrea Capa Salinas} 
\affiliation{
Materials Department, University of California Santa Barbara, Santa Barbara, California 93106, USA}

\author{\hspace{1mm}Stephen D. Wilson}
\affiliation{
Materials Department, University of California Santa Barbara, Santa Barbara, California 93106, USA}
 
\author{\hspace{1mm} Marc-Henri Julien}
 \affiliation{
  Universit\'e Grenoble Alpes, Laboratoire National des Champs Magnétiques Intenses, CNRS, Universit\'e de Toulouse, INSA Toulouse, European Magnetic Field Laboratory, 38042 Grenoble, France}

\author{\hspace{1mm}Samuele Sanna}
\affiliation{Dipartimento di Fisica e Astronomia, Universit\`a di Bologna, I-40127 Bologna, Italy}

\author{\hspace{1mm}Vesna F. Mitrovi{\'c}}
\email[]{corresponding author: vemi@brown.edu}
\affiliation{
Department of Physics, Brown University, Providence, Rhode Island 02912,   USA}
 \affiliation{
Brown Center for Theoretical Physics and Innovation, BCTPI, Brown University, Providence, RI 02912-1843, USA}

\date{\today}
\begin{abstract}
The interplay between superconductivity and charge-density wave (CDW) order, and its evolution with carrier density, is central to the physics of many quantum materials, notably high-$T_c$ cuprates and kagome metals. Hole-doped kagome compounds exhibit puzzling double-dome superconductivity and, as chemical substitution inevitably introduces quenched disorder, their properties remain poorly understood. Here, by leveraging the sensitivity of nuclear quadrupole resonance to local and static orderings, we uncover new features, primarily the incipient and fragmented CDW phases, in the charge landscape of CsV$_3$Sb$_{5-x}$Sn$_x$. Static CDW puddles are observed well above the transition temperature, a hallmark of pinning by defects. Their doping and temperature evolution indicate that, in the absence of disorder, the inverse Star-of-David $\pi$-shifted (\isdpi) CDW order would vanish near $x=0.12$,  between the two superconducting domes. This critical doping represents a hidden quantum critical point. Nevertheless, the \isdpi pattern persists well beyond previous reports, although its volume fraction is progressively reduced up to the critical doping at which it saturates. We establish that carrier doping promotes fragmentation of the \isdpi order, whereas randomness preserves the \isdpi patches. 
 \end{abstract}

\maketitle
Understanding the multifaceted nature of the relationship between charge density wave (CDW) and superconductivity (SC) is a major open question in condensed matter physics~\cite{Keimer2015}. 
Since both originate from effective attractive electron-electron interactions, they often compete but may coexist, or even enhance each other. Understanding their complex interplay can provide valuable insight into the mechanisms driving unconventional SC. The presence of disorder further complicates the physics to the extent that it can produce notable changes in the morphology of phase diagrams~(\cite{Alloul2009, Zhou2025} and refs. therein). 

The prototypical materials of such complex physics are cuprate superconductors, where despite over three decades of research, the interplay between SC and CDW has yet to be fully understood~\cite{Keimer2015}. An even greater complexity marks the family of vanadium-based $A$V$_3$Sb$_5$ \mbox{($A$ = K, Rb, Cs)} kagome superconductors~\cite{Wilson2024} due to the mixing of topologically nontrivial bands with structures that promote strong electronic correlations, \ie flat bands and Dirac cones~\cite{Li2018, Ghimire2020, Park2021, Denner2021, Lin2021, Christensen2021, Kang2022a, Yin2022, Wang2023a, Tazai2024}. In particular, \cvs has not fallen short of promise and exhibits a wide range of electronic instabilities, including 3Q CDW order~\cite{Li2023, Kautzsch2023, Guo2024, Bonfa2025} before reaching a SC ground state~\cite{Wilson2024}. The mechanism behind SC still remains a mystery, with proposals ranging from phonon-driven pairing to CDW and current-loop fluctuations~\cite{Wilson2024}. In order to understand how electronic instabilities govern the interplay between CDW and SC, one can tune the Fermi level by means of carrier doping and hydrostatic pressure. 

Studies on hope-doped \cvs have suggested that the CDW order disappears to give way to double-dome SC~\cite{Oey2022, Yang2022, Sur2023, Pokharel2025}. Doping inherently introduces defects into the lattice, which may disrupt CDW and/or alter SC. This has sparked debates about the intrinsic role of quenched disorder in shaping the CDW character even in archetypal materials~\cite{Zhou2025}. An important question is whether CDW is sensitive to disorder, which in turn modifies SC, or whether both are independently affected. In kagome systems, where topology, frustration, and strong interactions lead to exotic properties~\cite{Wilson2024}, the question becomes even more relevant. Untangling these effects from disorder is thus essential for understanding what fundamentally underpins the physics of kagome materials.  
 
\begin{figure*}[t]
   \centerline{\includegraphics[scale=0.61]{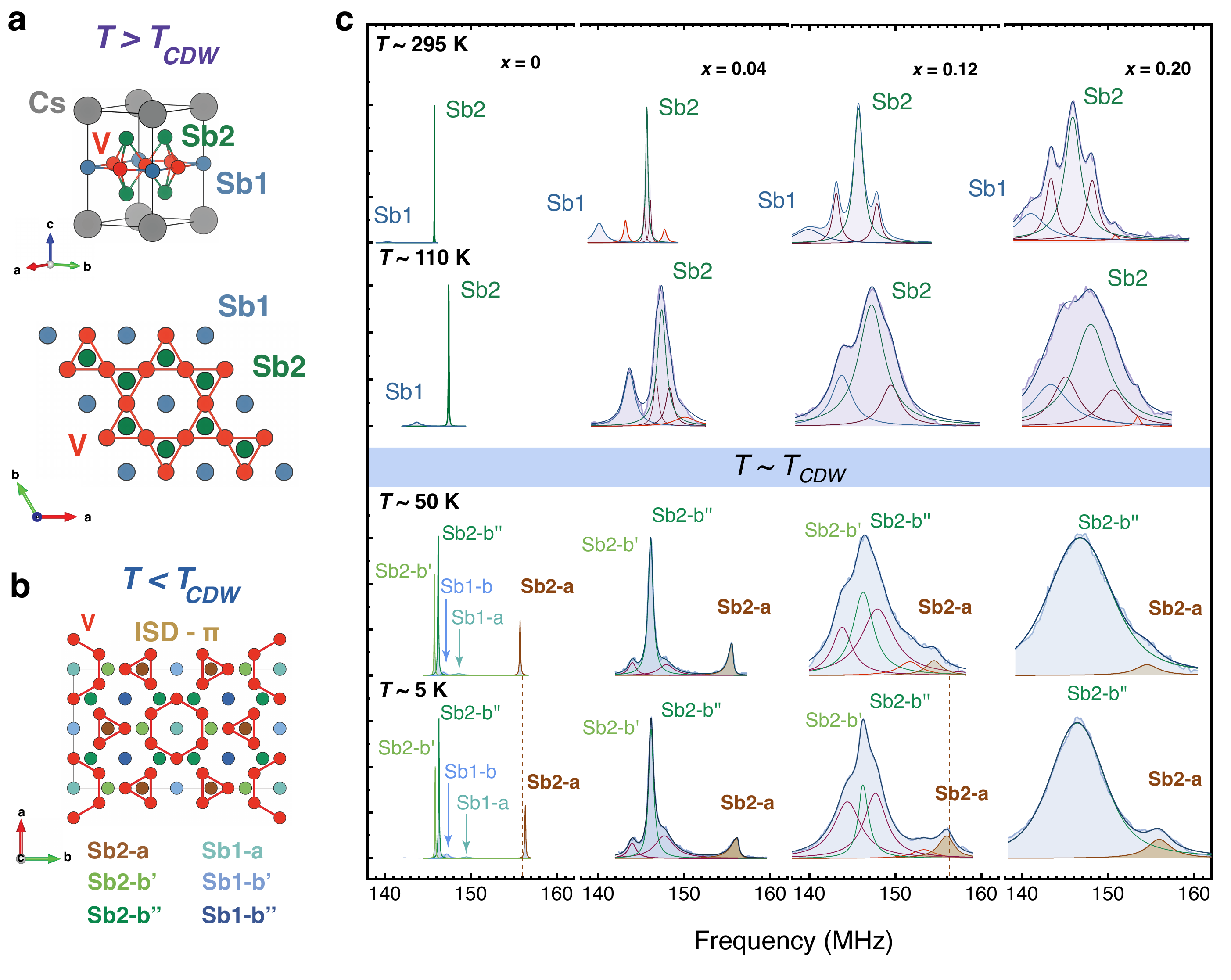}}
      \vspace*{-0.3cm}
    \caption{\textbf{Commensurate inverse Star-of-David CDW.} (\textbf{a}) Crystallographic structure of \cvs in the pristine phase. The kagome pattern is realized by the V atom in red. We distinguish two Sb sites: the basal Sb1 site, blue, and the apical Sb2 site, green. (\textbf{b}) Crystallographic structure of \cvs below $T_\text{CDW}$. The hexagonal lattice becomes orthorhombic with an inverse Star-of-David $\pi$-shifted modulation (\isdpi) (\textbf{c}) NQR \Sb spectra for the 5/2 $\rightarrow$ 3/2 transition as a function of temperature and Sn-doping $x$ in \cvsn. The solid curves represent Lorentzian-based models, while the data is the shaded color area. Even at high Sn-doping, charge ordering $T_\text{CDW}$ is unambiguously determined by the presence of the \mbox{Sb2-a} peak (brown curve) at approximately 156~MHz. For the undoped sample, below $T_\text{CDW}$, the whole sample is in the \isdpi phase. The Sb1-a and Sb1-b = Sb1-b' \& Sb1-b'' peaks are omitted for $x\ge 0.04$. The extra room temperature peaks belong to the Sb2 site based on site-occupancy analysis and dopant-induced neighboring-site effects. 
    }
    \label{fig:spectra_x}
       \vspace*{-0.4cm}
\end{figure*}

To identify the complex interaction between CDW, SC and disorder, as well as eliminate possible magnetic-field-induced artifacts,  we employ nuclear quadrupole resonance (NQR) on high-quality \cvsn powders~\cite{Ortiz2019}. NQR is a local probe whose response comes from the bulk and is sensitive to static effects. More importantly, since local magnetism (except the putative loop-current in the CDW~\cite{Wilson2024}) is absent in these compounds, NQR uniquely probes the charge density in the ground state. The power of this technique to determine the microscopic nature of the charge density ordering was demonstrated in RbV$_3$Sb$_5$~\cite{Frassineti2023}. Here, we apply the same methodology on the hole-doped \cvsn compounds and find that the lattice instability for $x=0$ is the inverse star of David, $\pi$-shifted along the $c$-axis (\isdpi), also known as staggered trihexagonal. We establish that this leading CDW instability (\isdpi) survives up to at least $x=0.35$, corresponding to the maximum of the second SC dome, and that the local CDW amplitude remains largely unchanged. The dominant effect of the dopant is to fragment the parent \isdpi order. 

Interestingly, at high temperatures, well above the transition temperature into the CDW order ($T_\text{CDW}$),  we detect nucleation of local static modulations of the charge density, {\it i.e.} CDW ``puddles.''  These puddles, present at all temperatures above $T_\text{CDW}$, arise from charge correlations once they are pinned by disorder. Moreover, the $T$-dependence of the correlations reveals a doping-tuned QCP that lies at the end of the first SC dome. The QCP indicates the critical doping at which the CDW would be suppressed in the absence of quenched disorder.
\newline

\noindent{\bf Nature of CDW in doped kagome lattice.} The kagome pattern in \cvs is realized by the vanadium atoms \mbox{(Fig.~\ref{fig:spectra_x}a \& b)} forming a quasi-2D structure. The Sn-substitution of similarly sized Sb-atoms, amounting to hole doping, is achieved with minimal steric hindrance~\cite{Oey2022}. In the pristine phase, above $T_\text{CDW}$, the system has two inequivalent antimony sites, Sb1 basal, center of V hexagons, and Sb2 apical, center of V triangles.  

The ${}^{121}$Sb nucleus (spin number $I=5\slash2$) has a finite quadrupole moment that couples to the electric field gradient (EFG) formed by the surrounding charge distribution. This coupling makes \Sb NQR a very sensitive probe of local charge arrangements. Because no ordered magnetism was detected in \cvs~\cite{Wilson2024}, the ${}^{121}$Sb  NQR observables can be directly related to the EFG. A particular peak in the NQR spectrum corresponds to a \Sb site in the unit cell, whereas the area under the peak (spectral area) is proportional to its occupancy in the unit cell (see Methods).

We first discuss the NQR spectra of the undoped sample, \cvs, shown in Fig.~\ref{fig:spectra_x}c. At room temperature, the two sites are easily discernible with the area ratio of (\mbox{Sb1 : Sb2} = \mbox{$1:4$}), in agreement with their respective site occupancies. When the temperature is reduced, $T<T_\text{CDW}$, the transition to the CDW phase is clearly identifiable by the appearance of the \mbox{Sb2-a} peak at higher frequencies, Fig.~\ref{fig:spectra_x}c. Following the procedure described in detail in Ref.~\cite{Frassineti2023}, we associate the appearance of the  \mbox{Sb2-a} peak with the emergence of the inverse Star-of-David (ISD) distortion pattern. As shown in Fig.~\ref{fig:spectra_x}b, the ISD pattern is realized on an orthorhombic unit cell and breaks the in-plane rotational symmetry. In addition, the \mbox{Sb2-b} peak splits into two spectral lines \mbox{(Fig.~\ref{fig:spectra_x}b \& c)}, which represents the hallmark of the $\pi$-shifted ISD~\cite{Frassineti2023}. The \isdpi pattern is marked by the staggering of layers of ISD that are $\pi$ shifted relative to neighboring layers along the $c$-axis within a 2x2x2 supercell. 

Next, we turn to NQR spectra analysis of the doped \cvsn for $T < T_\text{CDW}$. There are two features in the NQR spectra, the appearance of the Sb2-a site and the splitting of Sb2-b, that allow us to determine the microscopic nature of CDW. First, we detect the Sb2-a site for all doping levels, as depicted in light-brown shaded regions in Fig.~\ref{fig:spectra_x}c.  The presence of the  Sb2-a site  indicates that the ISD order persists up to the highest doping level investigated here $(x=0.35)$. Second, the splitting of Sb2-b into two additional sites that provides conclusive evidence for the $\pi$-shifted ISD is clearly perceptible only for doping levels up to $x=0.12$, as illustrated in Fig.~\ref{fig:spectra_x}c and Supplementary Note III. For $x\ge 0.20$, the splitting of the Sb2-b peak is not distinguishable due to excess spectral broadening. Nevertheless, the broadening does not impede us from determining whether interlayer correlations, \ie $\pi$ shift between neighboring layers, are present in the ISD-CDW phase.
  
\begin{figure}[ht]
    \centerline{\includegraphics[scale=0.32]{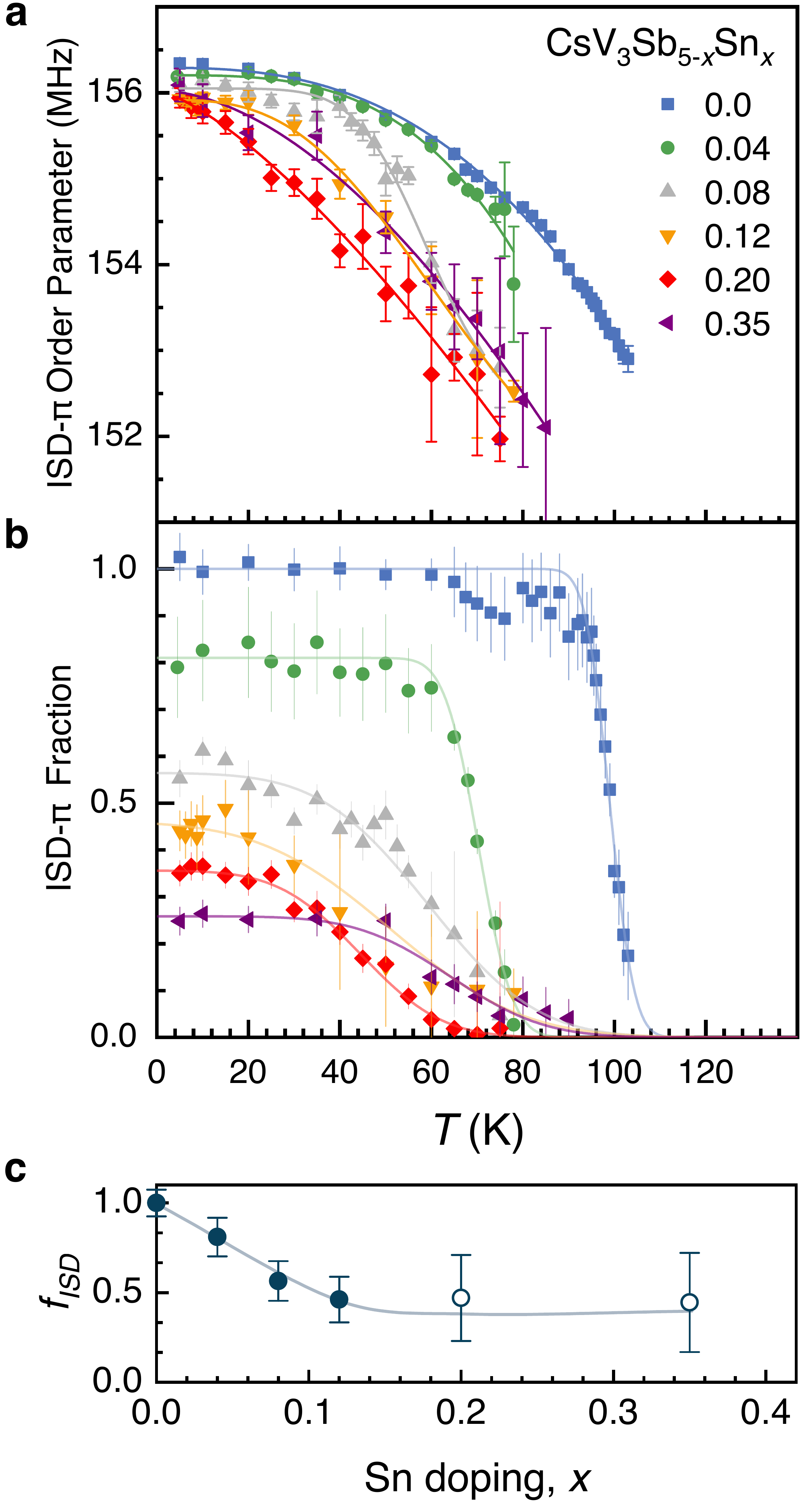}}
     \vspace*{-0.2cm}
    \caption{\textbf{CDW order parameter.} (\textbf{a}) The frequency of the Sb2-a site is proportional to the amplitude of the CDW modulation. The minimal change in the low temperature Sb2-a frequency shows that the CDW amplitude is almost unaltered by the dopant. 
    (\textbf{b}) The \isdpi volume fraction as a function of temperature.
    (\textbf{c}) The low-temperature volume-fraction ($f_\text{CDW}$) as a function of Sn-doping shows that the \isdpi survives in at least half of the sample. Open symbols are for data calculated by two techniques  (Supplementary Note II). Error bars represent two standard deviations and are not shown when smaller than the symbol.
    } 
    \label{fig:sb2a_rel_weigth}
     \vspace*{-0.3cm}
\end{figure}

Examining the spectral features enables us to extract the order parameter of the ISD-CDW, defined as the \mbox{Sb2-a} frequency, and its volume fraction, corresponding to the spectral area. In  \mbox{Fig.~\ref{fig:sb2a_rel_weigth}a} we plot the CDW order parameter as a function of temperature for the different doping levels. We observe that, for all doping levels, the order parameter reaches nearly the same frequency as the temperature approaches zero. Measurements in the undoped sample, where the spectral features are sharp,  demonstrate that the frequency difference between the two Sb2 sites is only compatible with \isdpi~\cite{Frassineti2023}. 
As doping increases, we find that the frequency difference between the two sites only slightly decreases. Therefore, as the frequency difference remains nearly unchanged with doping, we conclude that the \isdpi pattern survives. 

\begin{figure*}[t]
         \centerline{\includegraphics[scale=0.32]{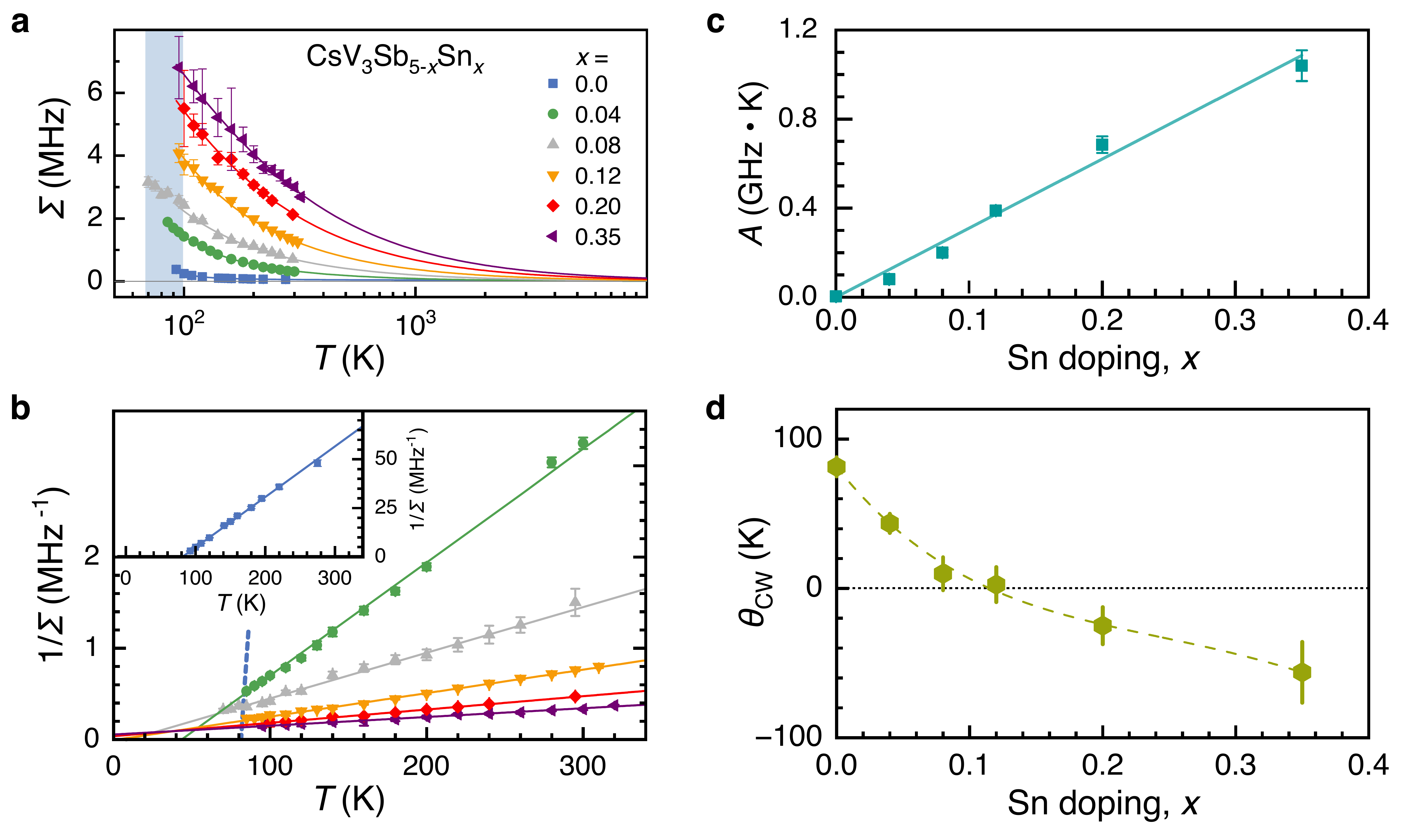}}
    \caption{\textbf{High-temperature charge correlations.} (\textbf{a}) Full-width-at-half-maximum ($\Sigma$) of the dominant Sb2 peak in the pristine phase $T>T_\text{CDW}$. The blue-shaded area represents the $T_\text{CDW}$ range for different dopings.
    (\textbf{b}) Linear fit to 1/$\Sigma$ demonstrates the Curie-Weiss (CW) behavior of the pristine-phase Sb2 linewidth and directly shows the intersection with the $x$-axis. The solid lines are CW fits. The inset is the undoped data. The dashed line is the CW fit of the undoped compound. Note that the undoped compound is a pure crystal as opposed to doped samples which are powders. 
    (\textbf{c}) The doping dependence of the CW parameter $A$ shows that Sn-doping is linearly proportional to the intrinsic CDW features.  
    (\textbf{d}) The CW temperature extracted from the linear fit shows a sign change in the vicinity of $x = 0.12$. The dashed green line is a guide to the eye. Error bars represent two standard deviations and are not shown when they are smaller than the symbol size.
    }
    \label{fig:Fig3_FWHM}
       \vspace*{-0.4cm}
\end{figure*}

To further probe the nature of the CDW as a function of doping, we examine the  volume fraction of \isdpi, determined as the ratio between the spectral area  under the \mbox{Sb2-a} peak and the total spectral area. We plot the $T$-dependence of the volume fraction and its value extrapolated at $T=0$ ($f_\text{ISD}$) in \mbox{Figs. \ref{fig:sb2a_rel_weigth}b and c}, respectively. The results show that $f_\text{ISD}$ is progressively reduced as a function of doping and saturates in the vicinity of  50~\%. In other words, the ISD order persists up to the highest doping level investigated but occupies no more than 50~\% of the sample. Thus, the system stabilizes in an inhomogeneous configuration in which the \isdpi order is fragmented. Additionally, recent XRD measurements deduced a drastic reduction in the correlation length at $x = 0.025$ for a 2x2 CDW in the plane~\cite{Kautzsch2023}. Together, these results affirm that the CDW is short-range, consistent with our fragmentation picture. Another key effect of dopant-induced fragmentation is the inevitable suppression of interlayer correlations, \ie the $\pi$-shift, without changing the nature of the CDW itself. Thus, the size of the fragment controls whether $\pi$-shift is present. 

An open question remains as to the exact nature of the additional phase (remaining 50~\% of volume fraction) that coexists with the \isdpi order at finite doping levels. With the aim of answering the question,  we simulate NQR spectra corresponding to different CDW patterns proposed in previous work~\cite{Zheng2022, Li2023,Feng2023, Kautzsch2023, Stier2024, Huai2025, Wu2025} and compare them with the measured spectra (Supplementary Note VII). These patterns include the $2a_0$~\cite{Zheng2022}, $4a_0$~\cite{Zheng2022, Li2023, Kautzsch2023}, $\frac{3}{8}a_0$~\cite{Stier2024}, $3a_0$~\cite{Huai2025} and incommensurate CDW~\cite{Feng2023, Kautzsch2023} on the hexagonal, $Fmmm$ and $C2m$ phases. Surprisingly, all of the potential charge arrangements explain equally well the spectral lineshape. Therefore, we cannot indisputably conclude which of the other proposed phases coexist with \isdpi at finite doping. Nonetheless, we establish that the presence of a layer of Star-of-David (SoD)~\cite{Deng2025} and/or alternating stacking of SoD and ISD layers~\cite{Wang2023} is incompatible with our findings (Supplementary Note VI). 
\newline

\noindent {\bf High-temperature charge correlations.} To better understand the CDW emergence, we proceed to analyze $^{121}$Sb NQR spectral details at higher temperatures \mbox{$(T > T_\text{CDW})$}. We find that NQR spectra feature linewidth broadening with a particular temperature evolution. Because in these compounds, \cvsn, there is no ordered magnetism~\cite{Wilson2024}, the broadening comes from the local charge distribution. From the linewidth, we extract the full-width-at-half-maximum ($\Sigma$) of the Sb2 peak and plot it in Fig.~\ref{fig:Fig3_FWHM}a as a function of temperature, $T>T_\text{CDW}$, for all doping levels. The $\Sigma$ follows a Curie-Weiss (CW) increase down to $T_\text{CDW}$. The solid line in \mbox{Fig.~\ref{fig:Fig3_FWHM}a} depicts a fit to the Curie-Weiss law
\begin{equation}
\label{eq:CW}
    \Sigma  \big(T\big) = \frac{A}{T-\theta_\text{CW}} + C,
\end{equation}
where $\theta_{CW}$ is the CW temperature, $C$  is the \mbox{infinite-$T$} linewidth, and $A$ is related to both the correlation length and maximum amplitude of the CDW modulation. 

The pre-transitional broadening of the NQR/NMR linewidth in CDW systems is a well-documented effect~\cite{CBerthier1978, Ghoshray2009, Wu2015, Vinograd2019, Feng2023}. 
Specifically, in the archetypal compound of charge ordering, NbSe$_2$, this broadening was associated with the formation of static short-range regions (puddles) of the incipient CDW~\cite{CBerthier1978}. The universality of the broadening in CDW compounds is illustrated in Fig.~S10. Furthermore, a similar effect has recently been observed in STM~(\cite{Liu2021} and refs. therein).

\begin{figure*}[ht!]
    \centerline{\includegraphics[scale=0.38]{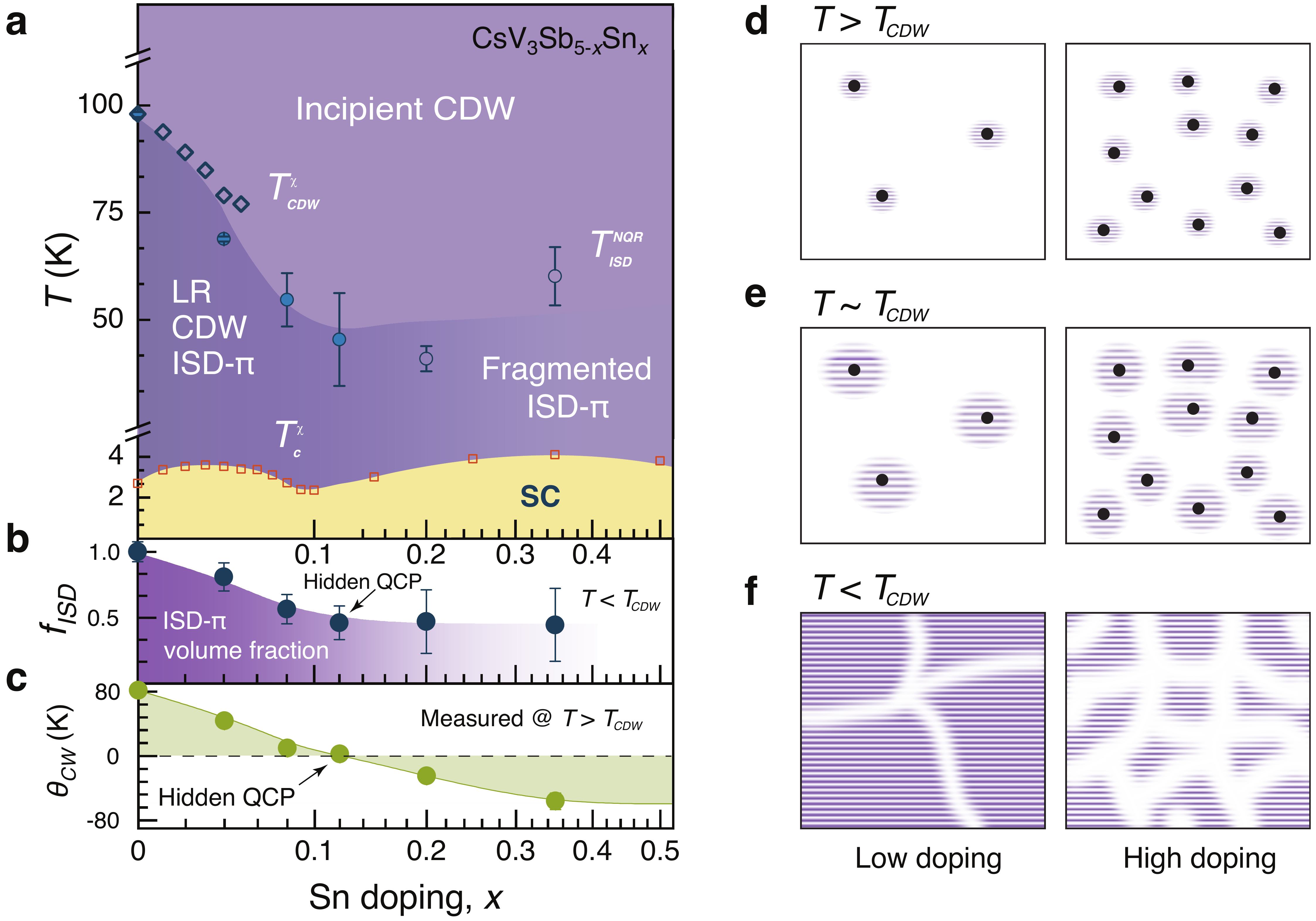}}
      \vspace*{-0.3cm}
    \caption{
    \textbf{Temperature phase diagram of \cvsn as a function of Sn-doping.}
       (\textbf{a}) Incipient CDW exists in the high temperature region, $T > T_\text{CDW}$. The \isdpi onset temperature is denoted by $T_\text{CDW}$, where LR denotes long-range order.
       The gradient represents crossover between the LR and fragmented order. 
       Open diamond and square symbols are magnetization $\chi$ measurements of the bulk CDW, dark blue, and SC, burned orange, transition temperatures, from~\cite{Oey2022}. (\textbf{b}) Volume fraction of \isdpi. Shown error bars reflect two standard deviations. (\textbf{c}) Doping-dependence of the CW temperature $\theta_\text{CW}$. The value $\theta_\text{CW} = 0$ indicates critical doping at which a putative hidden quantum critical point (QCP) lies. Error bars not shown for clarity. Solid lines serve as a guide to the eye.
    (\textbf{d-f}) Schematic of the $T$-dependent formation of CDW in the presence of quenched disorder for low and high doping. Black dots represent Sn-dopants and disorder, while purple lines represent charge modulations. (\textbf{d}) For $T>T_\text{CDW}$, Friedel oscillations occur around impurities to form incipient CDW. (\textbf{e}) On approaching $T_\text{CDW}$, the incipient CDW spreads. (\textbf{f}) For $T<T_\text{CDW}$, long-range CDW develops at low doping, while increased doping induces domain walls (white ribbon) that fragment the parent CDW.}
    \label{fig:phaseDiag}
      \vspace*{-0.5cm}
\end{figure*}

The emergence of puddles for $T>T_\text{CDW}$ appears naturally in realistic metals that inevitably contain impurities. Around them, Friedel oscillations occur, predominantly at the CDW wave vector if it exists. Quenched disorder can pin the fluctuations associated with these oscillations, inducing static components, \ie puddles of incipient CDW, that locally break the pristine translational symmetry.  This manifests itself in the NQR spectrum via linewidth broadening induced by local distortions that change the EFG around the nuclear active site. On approaching $T_\text{CDW}$, the volume fraction of the incipient CDW grows as illustrated in Fig.~\ref{fig:phaseDiag}d-f causing a further increase in the NQR linewidth as observed in Fig.~\ref{fig:Fig3_FWHM}.  

To extract the parameters of the CW law, we plot 1/$\Sigma(T)$ in Fig.~\ref{fig:Fig3_FWHM}b. Our fit finds that $C$ is zero within the error bar ($\pm$~0.1~MHz), suggesting that besides CDW pinning, the linewidth broadening is not strongly affected by Sn-doping. That is, doping does not appreciably distort the crystallographic structure so that it could introduce a significant distribution of EFG parameters. This finding ($C \approx 0$) is somewhat unusual because EFGs are quite sensitive to disorder and thus doped compounds would have broader lines in the high-temperature limit. Furthermore, in Fig.~\ref{fig:Fig3_FWHM}c, we plot the \mbox{parameter $A$} as a function of doping. More disorder means that there are more pinning sites at high temperatures, thus an increase of $A$. In the nominally undoped compound, $A$ remains finite but extremely small compared to that of the doped materials, which is an indication of the good quality of the undoped sample. 

The divergence in the CDW susceptibility is characterized by extracting the CW temperature, $\theta_\text{CW}$, as a function of doping (Fig.~\ref{fig:Fig3_FWHM}d). This intrinsic CDW susceptibility, following the CW $T$-dependence, is exclusively associated with electronic degrees of freedom. In other words, it is dissociated from randomness/disorder effects. The $\theta_\text{CW}$ exhibits a clear change of sign in the vicinity of $x=0.12$ doping and describes the growth of CDW fluctuations pinned by disorder, \ie puddles, as the system tends to order. Remarkably, while $\theta_\text{CW}$ changes sign at $x\approx 0.1$, Fig.~\ref{fig:Fig3_FWHM}d, for $T<T_\text{CDW}$ the CDW persists, as evidenced by both its order parameter and volume fraction (Fig.~\ref{fig:sb2a_rel_weigth}). This implies that CDW would be suppressed at $x\approx 0.12$ if no quenched disorder were present in the system. In the absence of quenched disorder, the sole effect of doping is to tune the Fermi level~\cite{Oey2022}. Therefore, the electronic ground state becomes unfavorable to CDW formation for doping levels exceeding $x\approx 0.12$, where $\theta_\text{CW}\le0$. However, the detection of short-range CDW up to $x=0.35$ establishes the crucial role of disorder, \ie intrinsic randomness of Sn-dopants, in stabilizing the CDW. In particular, the short-range CDW order survives as a result of fragmentation enabled by disorder. Thus, $x\approx 0.12$ represents a critical doping where a putative quantum critical point (QCP) hidden by disorder and possibly SC is harbored. This result is comparable to the one found in recent coherent phonon spectroscopy experiments, indicating that the critical doping (location of QCP) is independent of the energy scale of the probe~\cite{Kongruengkit2025, Huai2025}. Even for temperatures below $T_\text{CDW}$, there is a notable change of the CDW phase at this critical doping. More precisely, while we observe that $f_\text{ISD}$ saturates at 50~\%, the addition of disorder does not change the nature of CDW (Fig.~\ref{fig:sb2a_rel_weigth}). Furthermore, at the critical doping, there is an abrupt change in the width of the CDW transition (Fig.~S11). Therefore, beyond the critical doping, the salient order is the fragmented \isdpi.  

Next, we address the disordered character of the CDW phase transition. The identification of the incipient CDW order reveals that puddles of short-ranged CDW exist far above $T_\text{CDW}$. Our findings are in agreement with other techniques that detect dynamical responses in \cvs~\cite{Chen2022, Subires2023, Yang2023, Feng2023, Zhong2024, Frachet2024, Kongruengkit2025, Pokharel2025}. Specifically, recent x-ray diffusion scattering measurements concluded that the CDW in \cvs is of the order-disorder type with a critical growth of quasi-static domains~\cite{Subires2023}. Such quasi-static domains would be revealed in NQR spectral measurements only if they contain truly static components. Therefore, the presence of the incipient CDW, as identified by NQR, demonstrates that these fluctuations encompass static contributions. 
\newline

\noindent{\bf Hole-doping phase diagram. }
A comprehensive phase diagram for \cvsn as a function of hole doping in Fig.~\ref{fig:phaseDiag} maps out the boundaries of various charge orderings. Looking first at the \isdpi order parameter in Fig.~\ref{fig:sb2a_rel_weigth}, we extract the transition temperature of the CDW transition, $T_\text{CDW}$. What is more, the spectral area corresponding to the order parameter is used to establish \isdpi volume fraction, $f_\text{ISD}$ (Fig.~\ref{fig:sb2a_rel_weigth}b). By overlaying the volume fraction, the CW temperature and the SC transition temperature ($T_c$) as a function of doping, we reveal that the critical doping, corresponding to the hidden QCP ($x\approx 0.12$), coincides with the saturation point of the \isdpi volume fraction, and lies between the two SC domes. Moreover, we find that beyond this doping level ($x \gtrsim 0.12$), the subsisting order is the fragmented \isdpi. We point out that between the long-range and fragmented order, there must be a second-order phase transition in the vicinity of the critical doping that is most likely influenced by the nearby hidden QCP. At finite temperatures, fluctuations associated with the QCP could smear out such a phase transition, impeding its detection. Additionally, disorder-induced broadening of the phase transition can also obstruct its detection. 

Below, we discuss the effect of disorder on CDW formation. Above $T_\text{CDW}$, puddles nucleate around disorder sites. Specifically, from the linear growth of the CW constant $A$ with Sn-doping (Fig.~\ref{fig:Fig3_FWHM}c), we deduce that the puddles are trapped by the Sn-dopant sites that act as pinning centers, as illustrated in Fig.~\ref{fig:phaseDiag}d. For a fixed doping level, we observe that the linewidth increases significantly on approaching $T_{\rm CDW}$ as a consequence of the spreading of the puddles around the defect, as depicted in Fig.~\ref{fig:phaseDiag}e. When the puddles eventually overlap, they form the \isdpi order at $T \sim T_\text{CDW}$ (Fig.~\ref{fig:phaseDiag}f). This scenario neatly explains the CW temperature dependence of the zero-field CDW susceptibility, similar to the generalized susceptibility in x-ray measurements~\cite{Subires2023}. Below $T_\text{CDW}$, on the other hand, the effect of disorder is to fragment the long-range \isdpi order. Strictly speaking, each impurity slightly detunes the CDW modulations, which interfere with each other and generate domain walls (shown in white in Fig.~\ref{fig:phaseDiag}f). This fragments the parent \isdpi order that becomes short-ranged beyond the critical doping. There are two ways in which defects can induce the mismatch that ultimately leads to \isdpi fragmentation. For instance, at low-doping, for long-range coherence to be established, the CDW must escape pinning and move away from the defects~\cite{Liu2021}. In an alternative scenario, another CDW of unknown nature forms around impurities, while the parent CDW forms away from the defects~\cite{Yue2020}.

We now examine the role of Sn-doping in affecting intrinsic randomness in the kagome lattice. The CW behavior above $T_\text{CDW}$ allows us to study pure electronic properties corresponding to systems free of randomness. We find that at the same critical doping determined from the analysis of $\theta_\text{CW}$, there are significant changes in the properties that characterize the CDW phase below $T_\text{CDW}$ (Fig.~\ref{fig:phaseDiag}b). This implies that the Sn-dopants introduce intrinsic randomness and do not act as simple impurities. In other words, it is the randomness of the location distribution of the dopants that dictates the physics, even at high doping. This is consistent with recent XRD measurements on \cvsn affirming that there is no evidence of their macroscopic clustering~\cite{Kautzsch2023}. Nevertheless, even high-quality  samples, such as the ones used here, may appear locally heterogeneous on a microscopic length-scale due to random arrangements of dopant locations. 

In order to understand the intrinsic origin of the QCP, we contrast our findings with those from the pressure response. A recent NQR study uncovered a pressure-tuned QCP in the undoped \cvs~\cite{Feng2023}. For pressure levels higher than that corresponding to the QCP, the CDW is entirely suppressed~\cite{Chen2021, Li2022, Zheng2022, Feng2023, Stier2024}. This is in stark contrast to our finding that the ISD CDW survives at Sn-doping levels well above the QCP. What is more, at the QCP doping level, ISD volume fraction saturates. Therefore, quenched disorder plays a major role not only in enhancing the incipient CDW and causing nucleation around Sn-dopants, but it is also essential in sustaining the ISD distortion pattern via fragmentation at high doping. While the NQR pressure study observed a CW temperature dependence of the charge susceptibility, they found no evidence of charge ordering beyond the pressure-tuned QCP~\cite{Feng2023}. This confirms our conclusion that CDW persists due to randomness of dopant distribution that drives fragmentation. 

A further difference between pressure-tuned and doping-tuned QCPs is their location relative to the SC domes. The pressure-tuned QCP is at the pressure level that corresponds to the maximum of the second SC dome~\cite{Chen2021, Li2022, Zheng2022, Feng2023, Stier2024}, suggesting a possible link between the CDW and SC. However, the critical doping corresponding to QCP lies in the vicinity of the minimum of $T_c$ (Fig.~\ref{fig:phaseDiag}a), and thus, naively, the QCP weakens SC. The comparison may prove difficult as the QCP found by NQR is related to pure electronic effects in the absence of disorder, while $T_c$ was measured on samples with disorder. At a basic level, our results indicate that the CDW and SC exhibit a very complex relationship in these kagome systems. In addition, the effect of disorder may render the relationship even more intricate.  
\newline

\noindent{\bf Prospective.} 
In this work, we reveal that Sn-doping introduces holes that govern the electronic properties and intrinsic disorder due to their random locations in the lattice. As a consequence, the local \isdpi distortion pattern is identified at \mbox{$T \lesssim  T_{\rm CDW}$} markedly for all Sn-doping investigated, $x\le 0.35$.  However, its volume fraction and correlation length decrease as a consequence of randomness-induced fragmentation. For \mbox{$T \gg T_{\rm CDW}$}, short-ranged puddles are detected. Importantly, both incipient CDW and \isdpi persist as a function of doping beyond what was previously measured, up to at least the doping level corresponding to the maximum of the second SC dome~\cite{Oey2022, Kautzsch2023}. It was  proposed that even a small orbital-selective hole doping of the Sb bands would induce CDW renormalization, resulting in strong charge fluctuations and appearance of incommensurate, quasi-1D charge correlations preceded by the \isdpi suppression~\cite{Kautzsch2023}. Here, we determine that \isdpi persists, albeit fragmented (short-ranged) beyond the critical doping due to disorder, which explains why it has eluded detection until now~\cite{Oey2022, Kautzsch2023, Wilson2024}. Our findings are consistent with recent experimental observations of doping-induced short-range CDW~\cite{Yang2022, Zhu2024, Wu2025, Huai2025}.   

In these kagome systems, shifting the Fermi level to relevant points of the Brillouin zone is predicted to promote electronic orderings, including CDW order and unconventional SC~\cite{Wilson2024}. Previous theoretical studies have identified the V $d$ orbitals and apical Sb $p$ orbitals at the $M$-points as essential for the formation of saddle points, which are important for stabilizing different CDW phases~\cite{LaBollita2021, Tsirlin2022, Jeong2022, Ritz2023, Li2023Pi}. On the other hand, the basal Sb $p$ orbitals that form an electron-like pocket at the $\Gamma$-point have a negligible contribution to the saddle points~\cite{LaBollita2021, Tsirlin2022, Jeong2022, Ritz2023, Li2023Pi}. Applying hydrostatic pressure causes a movement of the apical Sb sites, while Sn-doping affects mostly the basal Sb sites~\cite{Oey2022}. Although pressure suppresses local CDW order at the maximum of the second dome~\cite{Zheng2022, Feng2023, Wilson2024}, here, we show that Sn-doping does not. Therefore, because Sn-doping primarily affects the Sb $p$ orbitals at the $\Gamma$-point~\cite{Oey2022}, we conclude that it is the Van Hove singularities at the $M$-points which are crucial for the stabilization of the CDW order, \isdpi, in these compounds.

The unveiling of the high-temperature charge correlations and local \isdpi order provides a new perspective on the complex interplay between the different types of electronic order and SC. It suggests that the nature of the SC state is different between the two domes separated by a QCP at the critical doping. Using local bulk probes, such as NQR, provides valuable information on the microscopic structure. Previous studies have found that doping suppresses the parent 3Q-CDW order~\cite{Nakayama2022, Yang2022,Kautzsch2023, Sur2023, Liu2023, Huai2024, Wilson2024, Kim2025, Wu2025}. Because no local-bulk probes in the absence of magnetic field were employed in the other doping-tuned kagome compounds, it would be useful to use NQR to determine if any local CDW persists and if disorder plays a role. Overall, our results emphasize the salient role of \textit{quenched disorder and tuning the Van Hove singularities near the} $M$-points in ascertaining what drives the staged electronic ordering of \cvsn. Ultimately, investigating the low-temperature state of $A$V$_3$Sb$_5$ kagome compounds as a function of doping through a local-bulk probe is an essential step to unravel the intricate role of disorder, topology, and charge correlations in stabilizing superconductivity.
\newpage

\noindent
\section*{Methods}
\label{sec:methods}
\noindent

\noindent{\bf Sample preparation}\\
The powder samples of \cvsn  with $x=$0, 0.04, 0.08, 0.12, 0.20, 0.35 are prepared in exactly the same way as in the ones used in refs.~\cite{Ortiz2019, Oey2022}. A single, high quality undoped \cvs crystal was used for the high temperature measurements ($T > T_\text{CDW}$).
\newline

\noindent{\bf Nuclear quadrupole resonance (NQR)} \\
Measurements on the powder samples were performed on a state-of-the-art, home-built phase-coherent, pulsed spectrometer at the Universit\`a di Parma. The powders were mounted on a specifically designed NQR coil in a sample holder that does not exert pressure. The measurement on the single crystal was performed on a different state-of-the-art, home-built phase-coherent, pulsed spectrometer at the LNCMI. The single crystal was mounted in a custom-built sample holder and placed under no external strain.

A typical spin echo technique is used to excite the nuclear spins similar to the one used in Ref.~\cite{Frassineti2023}. In NQR, the quadrupolar spins are polarized by the EFG and thus no applied magnetic field is needed. The $I=5/2$, \Sb nuclear spin have two NQR transitions. Even though higher transitions have a lower intensity, in naturally polarized thermal ensemble of spins, the higher frequency corresponds to a higher polarization factor, which justifies using the second transition. What is more, the peak separation in the second transition is twice the one in the lower transition, resulting in a better resolution.

Details on the curve fitting analysis are given in~Supplementary Note~I. As the spectral intensity is related to the static local environment on a timescale of the experiment, which in our case is in the $ms$ range, our measurements quantify primarily static contributions. 
\newline

\noindent{\bf Sb site-occupancy in CDW phase } \\\
Guided by density functional theory (DFT) calculations described in detail in \mbox{ref. \cite{Frassineti2023}}, we estimate the effect of the structural distortion on the \Sb sites and  reproduce the observed NQR spectrum. They are in a full agreement with presence of  \isdpi. 

The full analysis of the peak occupancy, including measurements on the first transition 1\slash2 $\rightarrow$ 3\slash2, indicate a perfect agreement of a single \isdpi modulation, at all temperatures below $T_\text{CDW}$, Fig.~\ref{fig:spectra_x}\textbf{c}. In Tab.~\ref{table:1} we give occupancy ratios of new antimony sites that develop.
\newline
\begin{table}[ht]
\centering
\resizebox{\columnwidth}{!}{%
\begin{tabular}{c c c} 
 \hline
Phase & Atoms & Multiplicity \\ [0.5ex] 
 \hline\hline
 Pristine & Sb1 & 1  \\
  Hexagonal - P6/mmm & Sb2 & 4 \\
\hline
2x2x2 CDW & Sb1-a / Sb1-b'/ Sb1-b'' & 2 / 2 / 4  \\ 
 \textit{Fmmm} & Sb2-a / Sb2-b'/ Sb2-b'' & 8 / 8 / 16  \\
 \small (ISD with $\pi$ shift) &  &  \\ [1ex] 
 \hline
\end{tabular}}
      \vspace*{-0.3cm}
\caption{The \Sb atomic occupations in the \cvs unit cell, with and without the \isdpi distortion. Taken from~\cite{Frassineti2023}}
\label{table:1}
      \vspace*{-0.3cm}
\end{table}

\noindent\textbf{Determination of $T_\text{CDW}$}\\ 
We model the area fraction in the prisine and CDW state (Fig.~\ref{fig:spectra_x}c) with a Gaussian distribution. The relative volume fraction is described by the Gauss error function (erf), which allows to quantitatively define the CDW onset (Fig.~\ref{fig:sb2a_rel_weigth}):
\begin{equation}
\label{eq:erf}
    1 - \text{erf}\left(\frac{T-T_0}{\sigma}\right),
\end{equation}
where $T_0$ is related to the center of the distribution, and $\sigma$ is related to the width of the transition temperatures. We define $T_\text{CDW}$ as the temperature at which the function defined in Eq.~\ref{eq:erf} reaches 80~\% of its maximum value. This allows for a consistent $T_\text{CDW}$ determination and quantification of the transition width, $\sigma$. 
\newline

\noindent{\bf \isdpi volume fraction calculation}\\
To quantify \isdpi, we start by evaluating the relative area between the Sb2-a peak and the full spectrum. By comparing the extracted Sb2-a fraction with the nominal \isdpi multiplicity expected for that site (see Table~\ref{table:1}), we determine the \isdpi volume fraction ($f_\text{ISD}$). The volume fraction for doping levels $x$ = 0.20, 0.35 is determined using two different methods, detailed in Supplementary
Note~I, in particular Fig.~S8.

\section*{Acknowledgments}
We thank J. M. Kosterlitz, A. Hui, D. Feldman, I. Vinograd and A.-A. Haghighirad for helpful discussions and H. Mayaffre for help with the measurements.
Work at Brown University was supported in part by the National Science Foundation grant No. DMR-1905532 and funds from Brown University and University of Bologna. Research at the LNCMI was supported in part by the Chateaubriand Fellowship of the Office for Science \& Technology of the Embassy of France in the USA. SDW and ACS gratefully acknowledge support via the UC Santa Barbara NSF Quantum Foundry funded through the Q-AMASE-i program under award DMR-1906325.  \\

\section*{Author contributions}
Crystals were synthesized and characterized by A.C.S. and S.D.W. Experiments were performed by G.A., I.K.N. and G.C. Analysis of experimental results was performed by I.K.N. and G.C. I.K.N., A.R., P.B., M.H.J., S.S. and V.F.M. developed and applied the general theoretical framework. I.K.N., A.R.,  M.H.J., S.S. and V.F.M. lead the experimental data interpretation.  V.F.M. and S.S. conceived and guided the project. All authors were involved in writing the paper.

\bibliography{bibliography}

@Article{Yang2023,
  author  = {Yang, Kunya and Xia, Wei and Mi, Xinrun and Zhang, Long and Gan, Yuhan and Wang, Aifeng and Chai, Yisheng and Zhou, Xiaoyuan and Yang, Xiaolong and Guo, Yanfeng and He, Mingquan},
  journal = {Physical Review B},
  title   = {Charge fluctuations above T {CDW} revealed by glasslike thermal transport in kagome metals A V 3 Sb 5 ( A = K , Rb , Cs )},
  year    = {2023},
  issn    = {2469-9950, 2469-9969},
  month   = may,
  number  = {18},
  pages   = {184506},
  volume  = {107},
  doi     = {10.1103/PhysRevB.107.184506},
}

@Misc{Pokharel2025,
  title = {Evolution of charge correlations in the hole-doped kagome superconductor ${\mathrm{CsV}}_{3\ensuremath{-}x}{\mathrm{Ti}}_{x}{\mathrm{Sb}}_{5}$},
  author = {Pokharel, Ganesh and Zhang, Canxun and Redekop, Evgeny and Ortiz, Brenden R. and Salinas, Andrea N. Capa and Schwarz, Sarah and Alvarado, Steven J. Gomez and Sarker, Suchismita and Young, Andrea F. and Wilson, Stephen D.},
  journal = {Phys. Rev. Mater.},
  volume = {9},
  issue = {9},
  pages = {094805},
  numpages = {8},
  year = {2025},
  month = {Sep},
  publisher = {American Physical Society},
  doi = {10.1103/gv69-1lyj},
  url = {https://link.aps.org/doi/10.1103/gv69-1lyj}
}

@Article{Zhou2025,
  author  = {Zhou, Rui and Vinograd, Igor and Mayaffre, Hadrien and Porras, Juan and Kim, Hun-Ho and Loew, Toshinao and Liu, Yiran and Le Tacon, Matthieu and Keimer, Bernhard and Julien, Marc-Henri},
  journal = {Physical Review Letters},
  title   = {Robust Charge Density Wave Correlations in Optimally Doped {YBa} 2 Cu 3 O y},
  year    = {2025},
  issn    = {0031-9007, 1079-7114},
  month   = sep,
  number  = {10},
  pages   = {106503},
  volume  = {135},
  doi     = {10.1103/8klr-4wc5},
}

@Article{Keimer2015,
  author  = {Keimer, B. and Kivelson, S. A. and Norman, M. R. and Uchida, S. and Zaanen, J.},
  journal = {Nature},
  title   = {From quantum matter to high-temperature superconductivity in copper oxides},
  year    = {2015},
  issn    = {0028-0836, 1476-4687},
  month   = feb,
  number  = {7538},
  pages   = {179--186},
  volume  = {518},
  doi     = {10.1038/nature14165},
  url     = {https://www.nature.com/articles/nature14165},
}

@Article{Li2023Pi,
  author  = {Li, Heqiu and Liu, Xiaoyu and Kim, Yong Baek and Kee, Hae-Young},
  journal = {Physical Review B},
  title   = {Origin of $\pi$ -shifted three-dimensional charge density waves in the kagom{\'{e}} metal A V 3 Sb 5 ( A = Cs , Rb , K )},
  year    = {2023},
  issn    = {2469-9950, 2469-9969},
  month   = aug,
  number  = {7},
  pages   = {075102},
  volume  = {108},
  doi     = {10.1103/PhysRevB.108.075102},
}

@Misc{Huai2025,
  author    = {Huai, Linwei and Wang, Zhuying and Rao, Huachen and Han, Yulei and Liu, Bo and Yu, Shuikang and Zhang, Yunmei and Zang, Ruiqing and Luan, Runqing and Peng, Shuting and Qiao, Zhenhua and Wang, Zhenyu and He, Junfeng and Wu, Tao and Chen, Xianhui},
  title     = {Electronic-correlation-assisted charge stripe order in a Kagome superconductor},
  year      = {2025},
  doi       = {10.48550/ARXIV.2509.17467},
  eprint    = {2509.17467},
  publisher = {{arXiv}},
    journal = {Preprint},
}

@Article{Tsirlin2022,
  author  = {Tsirlin, Alexander and Fertey, Pierre and Ortiz, Brenden R. and Klis, Berina and Merkl, Valentino and Dressel, Martin and Wilson, Stephen and Uykur, Ece},
  journal = {{SciPost} Physics},
  title   = {Role of Sb in the superconducting kagome metal {CsV}\$\_3\$Sb\$\_5\$ revealed by its anisotropic compression},
  year    = {2022},
  issn    = {2542-4653},
  month   = feb,
  number  = {2},
  pages   = {049},
  volume  = {12},
  doi     = {10.21468/SciPostPhys.12.2.049},
}

@Article{Jeong2022,
  author    = {Jeong, Min Yong and Yang, Hyeok-Jun and Kim, Hee Seung and Kim, Yong Baek and Lee, SungBin and Han, Myung Joon},
  journal   = {Phys. Rev. B},
  title     = {Crucial role of out-of-plane Sb $p$ orbitals in Van Hove singularity formation and electronic correlations in the superconducting kagome metal ${\mathrm{CsV}}_{3}{\mathrm{Sb}}_{5}$},
  year      = {2022},
  month     = jun,
  pages     = {235145},
  volume    = {105},
  doi       = {10.1103/PhysRevB.105.235145},
  issue     = {23},
  numpages  = {9},
  publisher = {American Physical Society},
}

@Article{Yang2022,
  author       = {Yang, Haitao and Huang, Zihao and Zhang, Yuhang and Zhao, Zhen and Shi, Jinan and Luo, Hailan and Zhao, Lin and Qian, Guojian and Tan, Hengxin and Hu, Bin and Zhu, Ke and Lu, Zouyouwei and Zhang, Hua and Sun, Jianping and Cheng, Jinguang and Shen, Chengmin and Lin, Xiao and Yan, Binghai and Zhou, Xingjiang and Wang, Ziqiang and Pennycook, Stephen J. and Chen, Hui and Dong, Xiaoli and Zhou, Wu and Gao, Hong-Jun},
  journal      = {Science Bulletin},
  title        = {Titanium doped kagome superconductor {CsV}3$-$Ti Sb5 and two distinct phases},
  year         = {2022},
  issn         = {2095-9273},
  month        = nov,
  number       = {21},
  pages        = {2176--2185},
  volume       = {67},
  doi          = {10.1016/j.scib.2022.10.015},
  shortjournal = {Science Bulletin},
  url          = {https://linkinghub.elsevier.com/retrieve/pii/S2095927322004753},
}

@Article{Sur2023,
  author       = {Sur, Yeahan and Kim, Kwang-Tak and Kim, Sukho and Kim, Kee Hoon},
  journal      = {Nature Communications},
  title        = {Optimized superconductivity in the vicinity of a nematic quantum critical point in the kagome superconductor Cs(V1-{xTix})3Sb5},
  year         = {2023},
  issn         = {2041-1723},
  month        = jul,
  number       = {1},
  pages        = {3899},
  volume       = {14},
  doi          = {10.1038/s41467-023-39495-1},
  shortjournal = {Nat Commun},
  url          = {https://www.nature.com/articles/s41467-023-39495-1},
}

@Article{Kim2025,
  author       = {Kim, Dongwook and Nam, H. W. and Sur, Yeahan and Kim, Kwang-Tak and Kim, Kee Hoon and Moon, S. J.},
  journal      = {Physical Review B},
  title        = {Doping evolution of the electronic response of the kagome metal Cs(V1$-${xTix})3Sb5},
  year         = {2025},
  issn         = {2469-9950, 2469-9969},
  month        = apr,
  note         = {Publisher: American Physical Society ({APS})},
  number       = {16},
  volume       = {111},
  doi          = {10.1103/physrevb.111.165151},
  rights       = {https://link.aps.org/licenses/aps-default-license},
  shortjournal = {Phys. Rev. B},
}

@Article{Liu2023,
  author       = {Liu, Yixuan and Wang, Yuan and Cai, Yongqing and Hao, Zhanyang and Ma, Xiao-Ming and Wang, Le and Liu, Cai and Chen, Jian and Zhou, Liang and Wang, Jinhua and Wang, Shanmin and He, Hongtao and Liu, Yi and Cui, Shengtao and Huang, Bing and Wang, Jianfeng and Chen, Chaoyu and Mei, Jia-Wei},
  journal      = {Physical Review Materials},
  title        = {Doping evolution of superconductivity, charge order, and band topology in hole-doped topological kagome superconductors Cs ( V 1 $-$ x Ti x ) 3 Sb 5},
  year         = {2023},
  issn         = {2475-9953},
  month        = jun,
  number       = {6},
  pages        = {064801},
  volume       = {7},
  doi          = {10.1103/PhysRevMaterials.7.064801},
  shortjournal = {Phys. Rev. Materials},
}

@Article{Kang2022a,
  author       = {Kang, Mingu and Fang, Shiang and Kim, Jeong-Kyu and Ortiz, Brenden R. and Ryu, Sae Hee and Kim, Jimin and Yoo, Jonggyu and Sangiovanni, Giorgio and Di Sante, Domenico and Park, Byeong-Gyu and Jozwiak, Chris and Bostwick, Aaron and Rotenberg, Eli and Kaxiras, Efthimios and Wilson, Stephen D. and Park, Jae-Hoon and Comin, Riccardo},
  journal      = {Nature Physics},
  title        = {Twofold van Hove singularity and origin of charge order in topological kagome superconductor {CsV}3Sb5},
  year         = {2022},
  issn         = {1745-2481},
  month        = mar,
  note         = {Publisher: Nature Publishing Group},
  number       = {3},
  pages        = {301--308},
  volume       = {18},
  doi          = {10.1038/s41567-021-01451-5},
  rights       = {2022 The Author(s), under exclusive licence to Springer Nature Limited},
  shortjournal = {Nat. Phys.},
  url          = {https://www.nature.com/articles/s41567-021-01451-5},
}

@Article{LaBollita2021,
  author       = {{LaBollita}, Harrison and Botana, Antia S.},
  journal      = {Physical Review B},
  title        = {Tuning the Van Hove singularities in A V 3 Sb 5 ( A = K , Rb , Cs ) via pressure and doping},
  year         = {2021},
  issn         = {2469-9950, 2469-9969},
  month        = nov,
  number       = {20},
  pages        = {205129},
  volume       = {104},
  doi          = {10.1103/PhysRevB.104.205129},
  shortjournal = {Phys. Rev. B},
}

@Article{Feng2023,
  author  = {Feng, X. Y. and Zhao, Z. and Luo, J. and Yang, J. and Fang, A. F. and Yang, H. T. and Gao, H. J. and Zhou, R. and Zheng, Guo-qing},
  journal = {npj Quantum Materials},
  title   = {Commensurate-to-incommensurate transition of charge-density-wave order and a possible quantum critical point in pressurized kagome metal {CsV3Sb5}},
  year    = {2023},
  issn    = {2397-4648},
  month   = may,
  number  = {1},
  pages   = {23},
  volume  = {8},
  doi     = {10.1038/s41535-023-00555-w},
  url     = {https://www.nature.com/articles/s41535-023-00555-w},
}

@Article{Zhu2024,
  author  = {Zhu, K. J. and Nie, L. P. and Lei, B. and Wang, M. J. and Sun, K. L. and Deng, Y. Z. and Tian, M. L. and Wu, T. and Chen, X. H.},
  journal = {Physical Review Research},
  title   = {Competitive superconductivity and charge density wave in {Mo}-doped kagome superconductor {Cs} {V} 3 {S} b 5},
  year    = {2024},
  issn    = {2643-1564},
  month   = jun,
  number  = {2},
  pages   = {023295},
  volume  = {6},
  doi     = {10.1103/PhysRevResearch.6.023295},
}

@Article{Huai2024,
  author  = {Huai, Linwei and Li, Hongyu and Han, Yulei and Luo, Yang and Peng, Shuting and Wei, Zhiyuan and Shen, Jianchang and Wang, Bingqian and Miao, Yu and Sun, Xiupeng and Ou, Zhipeng and Liu, Bo and Yu, Xiaoxiao and Xiang, Ziji and Kuang, Min-Quan and Qiao, Zhenhua and Chen, Xianhui and He, Junfeng},
  journal = {npj Quantum Materials},
  title   = {Two-dimensional phase diagram of the charge density wave in doped {CsV3Sb5}},
  year    = {2024},
  issn    = {2397-4648},
  month   = mar,
  number  = {1},
  pages   = {23},
  volume  = {9},
  doi     = {10.1038/s41535-024-00635-5},
  url     = {https://www.nature.com/articles/s41535-024-00635-5},
}

@Article{Nakayama2022,
  author  = {Nakayama, Kosuke and Li, Yongkai and Kato, Takemi and Liu, Min and Wang, Zhiwei and Takahashi, Takashi and Yao, Yugui and Sato, Takafumi},
  journal = {Physical Review X},
  title   = {Carrier {Injection} and {Manipulation} of {Charge}-{Density} {Wave} in {Kagome} {Superconductor} {CsV} 3 {Sb} 5},
  year    = {2022},
  issn    = {2160-3308},
  month   = jan,
  number  = {1},
  pages   = {011001},
  volume  = {12},
  doi     = {10.1103/PhysRevX.12.011001},
}

@Article{Chen2021,
  author  = {Chen, K.\hspace{0.167em}Y. and Wang, N.\hspace{0.167em}N. and Yin, Q.\hspace{0.167em}W. and Gu, Y.\hspace{0.167em}H. and Jiang, K. and Tu, Z.\hspace{0.167em}J. and Gong, C.\hspace{0.167em}S. and Uwatoko, Y. and Sun, J.\hspace{0.167em}P. and Lei, H.\hspace{0.167em}C. and Hu, J.\hspace{0.167em}P. and Cheng, J.-G.},
  journal = {Physical Review Letters},
  title   = {Double {Superconducting} {Dome} and {Triple} {Enhancement} of {T} c in the {Kagome} {Superconductor} {CsV} 3 {Sb} 5 under {High} {Pressure}},
  year    = {2021},
  issn    = {0031-9007, 1079-7114},
  month   = jun,
  number  = {24},
  pages   = {247001},
  volume  = {126},
  doi     = {10.1103/PhysRevLett.126.247001},
  ranking = {rank1},
}

@Article{Vinograd2019,
  author  = {Vinograd, I. and Zhou, R. and Mayaffre, H. and Kr{\"{a}}mer, S. and Liang, R. and Hardy, W. N. and Bonn, D. A. and Julien, M.-H.},
  journal = {Physical Review B},
  title   = {Nuclear magnetic resonance study of charge density waves under hydrostatic pressure in {YBa} 2 {Cu} 3 {O} y},
  year    = {2019},
  issn    = {2469-9950, 2469-9969},
  month   = sep,
  number  = {9},
  pages   = {094502},
  volume  = {100},
  doi     = {10.1103/PhysRevB.100.094502},
}

@Article{Subires2023,
  author  = {Subires, D. and Korshunov, A. and Said, A. H. and S{\'{a}}nchez, L. and Ortiz, Brenden R. and Wilson, Stephen D. and Bosak, A. and Blanco-Canosa, S.},
  journal = {Nature Communications},
  title   = {Order-disorder charge density wave instability in the kagome metal ({Cs},{Rb}){V3Sb5}},
  year    = {2023},
  issn    = {2041-1723},
  month   = feb,
  number  = {1},
  pages   = {1015},
  volume  = {14},
  doi     = {10.1038/s41467-023-36668-w},
  url     = {https://www.nature.com/articles/s41467-023-36668-w},
}

@Article{Christensen2021,
  author    = {Christensen, Morten H. and Birol, Turan and Andersen, Brian M. and Fernandes, Rafael M.},
  journal   = {Phys. Rev. B},
  title     = {Theory of the charge density wave in {$A$V}$_3${Sb}$_5$ kagome metals},
  year      = {2021},
  month     = dec,
  pages     = {214513},
  volume    = {104},
  doi       = {10.1103/PhysRevB.104.214513},
  issue     = {21},
  numpages  = {15},
  publisher = {American Physical Society},
  url       = {https://link.aps.org/doi/10.1103/PhysRevB.104.214513},
}

@Article{Lin2021,
  author    = {Lin, Yu-Ping and Nandkishore, Rahul M.},
  journal   = {Phys. Rev. B},
  title     = {Complex charge density waves at Van Hove singularity on hexagonal lattices: Haldane-model phase diagram and potential realization in the kagome metals {$A$V}$_3${Sb}$_5$ ({$A$} = {K}, {Rb}, {Cs})},
  year      = {2021},
  month     = {Jul},
  pages     = {045122},
  volume    = {104},
  doi       = {10.1103/PhysRevB.104.045122},
  issue     = {4},
  numpages  = {13},
  publisher = {American Physical Society},
  url       = {https://link.aps.org/doi/10.1103/PhysRevB.104.045122},
}

@Article{Denner2021,
  author    = {Denner, M. Michael and Thomale, Ronny and Neupert, Titus},
  journal   = {Phys. Rev. Lett.},
  title     = {Analysis of Charge Order in the Kagome Metal {$A$V}$_3${Sb}$_5$ ({$A$} = {K}, {Rb}, {Cs})},
  year      = {2021},
  month     = {Nov},
  pages     = {217601},
  volume    = {127},
  doi       = {10.1103/PhysRevLett.127.217601},
  issue     = {21},
  numpages  = {6},
  publisher = {American Physical Society},
  url       = {https://link.aps.org/doi/10.1103/PhysRevLett.127.217601},
}

@Article{Park2021,
  author    = {Park, Takamori and Ye, Mengxing and Balents, Leon},
  journal   = {Phys. Rev. B},
  title     = {Electronic instabilities of kagome metals: Saddle points and Landau theory},
  year      = {2021},
  month     = {Jul},
  pages     = {035142},
  volume    = {104},
  doi       = {10.1103/PhysRevB.104.035142},
  issue     = {3},
  numpages  = {20},
  publisher = {American Physical Society},
  url       = {https://link.aps.org/doi/10.1103/PhysRevB.104.035142},
}

@Article{Kautzsch2023,
  author  = {Kautzsch, Linus and Oey, Yuzki M. and Li, Hong and Ren, Zheng and Ortiz, Brenden R. and Pokharel, Ganesh and Seshadri, Ram and Ruff, Jacob and Kongruengkit, Terawit and Harter, John W. and Wang, Ziqiang and Zeljkovic, Ilija and Wilson, Stephen D.},
  journal = {npj Quantum Materials},
  title   = {Incommensurate charge-stripe correlations in the kagome superconductor {CsV}$_3${Sb}$_{5-x}${Sn}$_x$},
  year    = {2023},
  issn    = {2397-4648},
  number  = {1},
  pages   = {37},
  volume  = {8},
  doi     = {10.1038/s41535-023-00570-x},
  refid   = {Kautzsch2023},
  url     = {https://doi.org/10.1038/s41535-023-00570-x},
}

@Article{Ritz2023,
  author    = {Ritz, Ethan T. and Fernandes, Rafael M. and Birol, Turan},
  journal   = {Phys. Rev. B},
  title     = {Impact of {Sb} degrees of freedom on the charge density wave phase diagram of the kagome metal {CsV}$_{3}${Sb}$_{5}$},
  year      = {2023},
  month     = may,
  pages     = {205131},
  volume    = {107},
  doi       = {10.1103/PhysRevB.107.205131},
  issue     = {20},
  numpages  = {12},
  publisher = {American Physical Society},
  url       = {https://link.aps.org/doi/10.1103/PhysRevB.107.205131},
}

@Article{Guo2024,
  author  = {Guo, Chunyu and Wagner, Glenn and Putzke, Carsten and Chen, Dong and Wang, Kaize and Zhang, Ling and Gutierrez-Amigo, Martin and Errea, Ion and Vergniory, Maia G. and Felser, Claudia and Fischer, Mark H. and Neupert, Titus and Moll, Philip J. W.},
  journal = {Nature Physics},
  title   = {Correlated order at the tipping point in the kagome metal {CsV}$_3${Sb}$_5$},
  year    = {2024},
  issn    = {1745-2473, 1745-2481},
  month   = apr,
  number  = {4},
  pages   = {579--584},
  volume  = {20},
  doi     = {10.1038/s41567-023-02374-z},
  url     = {https://www.nature.com/articles/s41567-023-02374-z},
}

@Article{Bonfa2025,
  author    = {Bonf{\`{a}}, Pietro and Pratt, Francis and Valenti, Diego and Onuorah, Ifeanyi John and Kataria, Anshu and Baker, Peter J. and Cottrell, Stephen and Salinas, Andrea Capa and Wilson, Stephen D. and Guguchia, Zurab and Sanna, Samuele},
  journal   = {Phys. Rev. Res.},
  title     = {Unveiling the nature of electronic transitions in RbV${}_{3}$Sb${}_{5}$ with avoided level crossing $\ensuremath{\mu}$SR},
  year      = {2025},
  doi       = {10.1103/bvgk-q2qn},
  publisher = {American Physical Society},
}

@Article{Wang2023a,
  author  = {Wang, Yaojia and Wu, Heng and {McCandless}, Gregory T. and Chan, Julia Y. and Ali, Mazhar N.},
  journal = {Nature Reviews Physics},
  title   = {Quantum states and intertwining phases in kagome materials},
  year    = {2023},
  issn    = {2522-5820},
  month   = sep,
  number  = {11},
  pages   = {635--658},
  volume  = {5},
  doi     = {10.1038/s42254-023-00635-7},
  url     = {https://www.nature.com/articles/s42254-023-00635-7},
}

@Article{Yin2022,
  author  = {Yin, Jia-Xin and Lian, Biao and Hasan, M. Zahid},
  journal = {Nature},
  title   = {Topological kagome magnets and superconductors},
  year    = {2022},
  issn    = {0028-0836, 1476-4687},
  month   = dec,
  number  = {7941},
  pages   = {647--657},
  volume  = {612},
  doi     = {10.1038/s41586-022-05516-0},
  url     = {https://www.nature.com/articles/s41586-022-05516-0},
}

@Misc{Kongruengkit2025,
  author    = {Kongruengkit, Terawit and Salinas, Andrea N. Capa and Pokharel, Ganesh and Ortiz, Brenden R. and Wilson, Stephen D. and Harter, John W.},
  month     = aug,
  title     = {Persistence of charge density wave fluctuations in the absence of long-range order in a hole-doped kagome metal},
  year      = {2025},
  doi       = {10.48550/arXiv.2508.13290},
  eprint    = {2508.13290},
  number    = {{arXiv}:2508.13290},
  publisher = {{arXiv}},
}

@Article{Tazai2024,
  author  = {Tazai, Rina and Yamakawa, Youichi and Kontani, Hiroshi},
  journal = {Proceedings of the National Academy of Sciences},
  title   = {Drastic magnetic-field-induced chiral current order and emergent current-bond-field interplay in kagome metals},
  year    = {2024},
  issn    = {0027-8424, 1091-6490},
  month   = jan,
  number  = {3},
  pages   = {e2303476121},
  volume  = {121},
  doi     = {10.1073/pnas.2303476121},
  url     = {https://pnas.org/doi/10.1073/pnas.2303476121},
}

@Article{Zheng2022,
  author  = {Zheng, Lixuan and Wu, Zhimian and Yang, Ye and Nie, Linpeng and Shan, Min and Sun, Kuanglv and Song, Dianwu and Yu, Fanghang and Li, Jian and Zhao, Dan and Li, Shunjiao and Kang, Baolei and Zhou, Yanbing and Liu, Kai and Xiang, Ziji and Ying, Jianjun and Wang, Zhenyu and Wu, Tao and Chen, Xianhui},
  journal = {Nature},
  title   = {Emergent charge order in pressurized kagome superconductor {CsV}$_3${Sb}$_5$},
  year    = {2022},
  issn    = {0028-0836, 1476-4687},
  month   = nov,
  number  = {7937},
  pages   = {682--687},
  volume  = {611},
  doi     = {10.1038/s41586-022-05351-3},
}

@Article{Frassineti2023,
  author    = {Frassineti, Jonathan and Bonf\`a, Pietro and Allodi, Giuseppe and Garcia, Erick and Cong, Rong and Ortiz, Brenden R. and Wilson, Stephen D. and De Renzi, Roberto and Mitrovi\ifmmode \acute{c}\else \'{c}\fi{}, Vesna F. and Sanna, Samuele},
  journal   = {Phys. Rev. Res.},
  title     = {Microscopic nature of the charge-density wave in the kagome superconductor {RbV}$_{3}${Sb}$_{5}$},
  year      = {2023},
  month     = feb,
  pages     = {L012017},
  volume    = {5},
  doi       = {10.1103/PhysRevResearch.5.L012017},
  issue     = {1},
  numpages  = {7},
  publisher = {American Physical Society},
  url       = {https://link.aps.org/doi/10.1103/PhysRevResearch.5.L012017},
}

@Article{Wu2015,
  author       = {Wu, Tao and Mayaffre, Hadrien and Kr{\"{a}}mer, Steffen and Horvati{\'{c}}, Mladen and Berthier, Claude and Hardy, W.N. and Liang, Ruixing and Bonn, D.A. and Julien, Marc-Henri},
  journal      = {Nature Communications},
  title        = {Incipient charge order observed by {NMR} in the normal state of {YBa}2Cu3Oy},
  issn         = {2041-1723},
  number       = {1},
  pages        = {6438},
  volume       = {6},
  date         = {2015-03-09},
  doi          = {10.1038/ncomms7438},
  langid       = {english},
  shortjournal = {Nat Commun},
year = {2015},
  url          = {https://www.nature.com/articles/ncomms7438},
}

@Article{CBerthier1978,
  author       = {Claude, Berthier and D, Jerome and P, Molinie},
  journal      = {Journal of Physics C: Solid State Physics},
  title        = {{NMR} study on a 2H-{NbSe}$_{\textrm{2}}$ single crystal: A microscopic investigation of the charge density waves state},
  issn         = {0022-3719},
  number       = {4},
  pages        = {797--814},
  volume       = {11},
  date         = {1978-02-28},
  doi          = {10.1088/0022-3719/11/4/024},
year = {1978},
  shortjournal = {J. Phys. C: Solid State Phys.},
  shorttitle   = {{NMR} study on a 2H-{NbSe}$_{\textrm{2}}$ single crystal},
}

@Article{Liu2021,
  author  = {Liu, Limin and Zhu, Changjiang and Liu, Z.\hspace{0.167em}Y. and Deng, Hanbin and Zhou, X.\hspace{0.167em}B. and Li, Yuan and Sun, Yingkai and Huang, Xiong and Li, Shuaishuai and Du, Xin and Wang, Zheng and Guan, Tong and Mao, Hanqing and Sui, Y. and Wu, Rui and Yin, Jia-Xin and Cheng, J.-G. and Pan, Shuheng\hspace{0.167em}H.},
  journal = {Physical Review Letters},
  title   = {Thermal Dynamics of Charge Density Wave Pinning in {ZrTe} 3},
  year    = {2021},
  issn    = {0031-9007, 1079-7114},
  month   = jun,
  number  = {25},
  pages   = {256401},
  volume  = {126},
  doi     = {10.1103/PhysRevLett.126.256401},
}

@Article{Ghoshray2009,
  author  = {Ghoshray, K and Pahari, B and Ghoshray, A and Eremenko, V V and Sirenko, V A and Suits, B H},
  journal = {Journal of Physics: Condensed Matter},
  title   = {$^{\textrm{93}}$ Nb {NMR} study of the charge density wave state in {NbSe}$_{\textrm{2}}$},
  year    = {2009},
  issn    = {0953-8984, 1361-648X},
  month   = apr,
  number  = {15},
  pages   = {155701},
  volume  = {21},
  doi     = {10.1088/0953-8984/21/15/155701},
}

@Article{Frachet2024,
  author  = {Frachet, Mehdi and Wang, Liran and Xia, Wei and Guo, Yanfeng and He, Mingquan and Maraytta, Nour and Heid, Rolf and Haghighirad, Amir-Abbas and Merz, Michael and Meingast, Christoph and Hardy, Fr{\'{e}}d{\'{e}}ric},
  journal = {Physical Review Letters},
  title   = {Colossal c -Axis Response and Lack of Rotational Symmetry Breaking within the Kagome Planes of the {CsV} 3 Sb 5 Superconductor},
  year    = {2024},
  issn    = {0031-9007, 1079-7114},
  month   = may,
  number  = {18},
  pages   = {186001},
  volume  = {132},
  doi     = {10.1103/PhysRevLett.132.186001},
}

@Article{Li2022,
  author       = {Li, Haoxiang and Fabbris, G. and Said, A. H. and Sun, J. P. and Jiang, Yu-Xiao and Yin, J.-X. and Pai, Yun-Yi and Yoon, Sangmoon and Lupini, Andrew R. and Nelson, C. S. and Yin, Q. W. and Gong, C. S. and Tu, Z. J. and Lei, H. C. and Cheng, J.-G. and Hasan, M. Z. and Wang, Ziqiang and Yan, Binghai and Thomale, R. and Lee, H. N. and Miao, H.},
  journal      = {Nature Communications},
  title        = {Discovery of conjoined charge density waves in the kagome superconductor {CsV}$_3${Sb}$_5$},
  year         = {2022},
  issn         = {2041-1723},
  month        = oct,
  number       = {1},
  pages        = {6348},
  volume       = {13},
  doi          = {10.1038/s41467-022-33995-2},
  shortjournal = {Nat Commun},
  url          = {https://www.nature.com/articles/s41467-022-33995-2},
}

@Misc{Deng2025,
  author    = {Deng, Qinwen and Tan, Hengxin and Ortiz, Brenden R. and Salinas, Andrea Capa and Wilson, Stephen D. and Yan, Binghai and Wu, Liang},
  month     = mar,
  title     = {Coherent Phonon Pairs and Rotational Symmetry Breaking of Charge Density Wave Order in the Kagome Metal {CsV}$_3${Sb}$_5$},
  year      = {2025},
  doi       = {10.48550/arXiv.2503.07442},
  eprint    = {2503.07442},
  number    = {{arXiv}:2503.07442},
  publisher = {{arXiv}},
  journal = {Preprint},
}

@Article{Stier2024,
  author       = {Stier, F. and Haghighirad, A.-A. and Garbarino, G. and Mishra, S. and Stilkerich, N. and Chen, D. and Shekhar, C. and Lacmann, T. and Felser, C. and Ritschel, T. and Geck, J. and Le Tacon, M.},
  journal      = {Physical Review Letters},
  title        = {Pressure-Dependent Electronic Superlattice in the Kagome Superconductor {CsV}$_3${Sb}$_5$},
  year         = {2024},
  issn         = {0031-9007, 1079-7114},
  month        = dec,
  number       = {23},
  pages        = {236503},
  volume       = {133},
  doi          = {10.1103/PhysRevLett.133.236503},
  langid       = {english},
  shortjournal = {Phys. Rev. Lett.},
}

@Article{Li2023,
  author  = {Li, Hong and Zhao, He and Ortiz, Brenden R. and Oey, Yuzki and Wang, Ziqiang and Wilson, Stephen D. and Željkovi\'c, Ilija},
  journal = {Nature Physics},
  title   = {Unidirectional coherent quasiparticles in the high-temperature rotational symmetry broken phase of {$A$}{V}$_3${Sb}$_5$ kagome superconductors},
  year    = {2023},
  issn    = {1745-2481},
  number  = {5},
  pages   = {637--643},
  volume  = {19},
  doi     = {10.1038/s41567-022-01932-1},
  refid   = {Li2023},
  url     = {https://doi.org/10.1038/s41567-022-01932-1},
}

@article {Li2018,
	author = {Li, Zhi and Zhuang, Jincheng and Wang, Li and Feng, Haifeng and Gao, Qian and Xu, Xun and Hao, Weichang and Wang, Xiaolin and Zhang, Chao and Wu, Kehui and Dou, Shi Xue and Chen, Lan and Hu, Zhenpeng and Du, Yi},
	title = {{Realization of flat band with possible nontrivial topology in electronic Kagome lattice}},
	volume = {4},
	number = {11},
	year = {2018},
	doi = {10.1126/sciadv.aau4511},
	publisher = {American Association for the Advancement of Science},
	url = {https://advances.sciencemag.org/content/4/11/eaau4511},
	journal = {Science Advances}
}

@article {Ghimire2020,
	author = {Ghimire, N.J. and Mazin, I.I.},
	title = {{Topology and correlations on the kagome lattice}},
	volume = {19},
	year = {2020},
	pages = {137-138},
	doi = {https://doi.org/10.1038/s41563-019-0589-8},
	url = {https://www.nature.com/articles/s41563-019-0589-8#citeas},
	journal = {Nature Materials}
}

@Article{Wilson2024,
  author  = {Wilson, Stephen D. and Ortiz, Brenden R.},
  journal = {Nature Reviews Materials},
  title   = {{$A$V}$_3${Sb}$_5$ kagome superconductors},
  year    = {2024},
  issn    = {2058-8437},
  number  = {6},
  pages   = {420--432},
  volume  = {9},
  doi     = {10.1038/s41578-024-00677-y},
  refid   = {Wilson2024},
  url     = {https://doi.org/10.1038/s41578-024-00677-y},
}

@Article{Ortiz2019,
  author    = {Ortiz, Brenden R. and Gomes, L\'{\i}dia C. and Morey, Jennifer R. and Winiarski, Michal and Bordelon, Mitchell and Mangum, John S. and Oswald, Iain W. H. and Rodriguez-Rivera, Jose A. and Neilson, James R. and Wilson, Stephen D. and Ertekin, Elif and McQueen, Tyrel M. and Toberer, Eric S.},
  journal   = {Phys. Rev. Materials},
  title     = {{New kagome prototype materials: discovery of ${\mathrm{KV}}_{3}{\mathrm{Sb}}_{5},{\mathrm{RbV}}_{3}{\mathrm{Sb}}_{5}$, and ${\mathrm{CsV}}_{3}{\mathrm{Sb}}_{5}$}},
  year      = {2019},
  month     = sep,
  pages     = {094407},
  volume    = {3},
  doi       = {10.1103/PhysRevMaterials.3.094407},
  issue     = {9},
  numpages  = {9},
  publisher = {American Physical Society},
  url       = {https://link.aps.org/doi/10.1103/PhysRevMaterials.3.094407},
}

@Article{Zhong2024,
  author  = {Zhong, Yigui and Suzuki, Takeshi and Liu, Hongxiong and Liu, Kecheng and Nie, Zhengwei and Shi, Youguo and Meng, Sheng and Lv, Baiqing and Ding, Hong and Kanai, Teruto and Itatani, Jiro and Shin, Shik and Okazaki, Kozo},
  journal = {Physical Review Research},
  title   = {Unveiling van Hove singularity modulation and fluctuated charge order in kagome superconductor Cs V 3 S b 5 via time-resolved {ARPES}},
  year    = {2024},
  issn    = {2643-1564},
  month   = dec,
  number  = {4},
  pages   = {043328},
  volume  = {6},
  doi     = {10.1103/PhysRevResearch.6.043328},
}

@Article{Chen2022,
  author  = {Chen, Q. and Chen, D. and Schnelle, W. and Felser, C. and Gaulin, B.\hspace{0.167em}D.},
  journal = {Physical Review Letters},
  title   = {Charge Density Wave Order and Fluctuations above T {CDW} and below Superconducting T c in the Kagome Metal {CsV} 3 Sb 5},
  year    = {2022},
  issn    = {0031-9007, 1079-7114},
  month   = jul,
  number  = {5},
  pages   = {056401},
  volume  = {129},
  doi     = {10.1103/PhysRevLett.129.056401},
}

@Article{Alloul2009,
  author  = {Alloul, H. and Bobroff, J. and Gabay, M. and Hirschfeld, P. J.},
  journal = {Reviews of Modern Physics},
  title   = {Defects in correlated metals and superconductors},
  year    = {2009},
  issn    = {0034-6861, 1539-0756},
  month   = jan,
  number  = {1},
  pages   = {45--108},
  volume  = {81},
  doi     = {10.1103/RevModPhys.81.45},
}

@Article{Oey2022,
  author    = {Oey, Yuzki M. and Ortiz, Brenden R. and Kaboudvand, Farnaz and Frassineti, Jonathan and Garcia, Erick and Cong, Rong and Sanna, Samuele and Mitrovi\'c, Vesna F. and Seshadri, Ram and Wilson, Stephen D.},
  journal   = {Phys. Rev. Materials},
  title     = {{Fermi level tuning and double-dome superconductivity in the kagome metal ${\mathrm{CsV}}_{3}{\mathrm{Sb}}_{5\ensuremath{-}x}{\mathrm{Sn}}_{x}$}},
  year      = {2022},
  month     = apr,
  pages     = {L041801},
  volume    = {6},
  doi       = {10.1103/PhysRevMaterials.6.L041801},
  issue     = {4},
  numpages  = {6},
  publisher = {American Physical Society},
  url       = {https://link.aps.org/doi/10.1103/PhysRevMaterials.6.L041801},
}

@Article{Yue2020,
  author  = {Yue, Li and Xue, Shangjie and Li, Jiarui and Hu, Wen and Barbour, Andi and Zheng, Feipeng and Wang, Lichen and Feng, Ji and Wilkins, Stuart B. and Mazzoli, Claudio and Comin, Riccardo and Li, Yuan},
  journal = {Nature Communications},
  title   = {Distinction between pristine and disorder-perturbed charge density waves in {ZrTe}3},
  year    = {2020},
  issn    = {2041-1723},
  month   = jan,
  number  = {1},
  pages   = {98},
  volume  = {11},
  doi     = {10.1038/s41467-019-13813-y},
  url     = {https://www.nature.com/articles/s41467-019-13813-y},
}

@Article{Wu2025,
  author  = {Wu, Zhimian and Sun, Kuanglv and Li, Hongyu and Nie, Linpeng and Rao, Huachen and Zhao, Dan and Xiang, Ziji and Ying, Jianjun and Wang, Zhenyu and Wu, Tao and Chen, Xianhui},
  journal = {Physical Review B},
  title   = {Competitive charge density waves in the doped kagome superconductor {CsV} 3 $-$ x Ti x Sb 5},
  year    = {2025},
  issn    = {2469-9950, 2469-9969},
  month   = oct,
  number  = {14},
  pages   = {144512},
  volume  = {112},
  doi     = {10.1103/ls5k-8q3d},
}

@article{Wang2023,
  title = {Structure of the kagome superconductor ${\mathrm{CsV}}_{3}{\mathrm{Sb}}_{5}$ in the charge density wave state},
  author = {Wang, Yuxin and Wu, Tao and Li, Zheng and Jiang, Kun and Hu, Jiangping},
  journal = {Phys. Rev. B},
  volume = {107},
  issue = {18},
  pages = {184106},
  numpages = {12},
  year = {2023},
  month = {May},
  publisher = {American Physical Society},
  doi = {10.1103/PhysRevB.107.184106},
  url = {https://link.aps.org/doi/10.1103/PhysRevB.107.184106}
}
\end{document}